\documentstyle[12pt]{article}
\def\beq{\begin{equation}}
\def\eeq{\end{equation}}
\def\bea{\begin{eqnarray}}
\def\eea{\end{eqnarray}}
\def\bq{\begin{quote}}
\def\eq{\end{quote}}

\def\PLB{{\it Phys. Lett.} }
\def\PRL{{\it Phys.Rev.Lett.} }
\def\NP{{\it Nucl.Phys.} }
\def\PR{{\it Phys.Rev.} }

\def\gappeq{\mathrel{\rlap {\raise.5ex\hbox{$>$}}
{\lower.5ex\hbox{$\sim$}}}}

\def\lappeq{\mathrel{\rlap{\raise.5ex\hbox{$<$}}
{\lower.5ex\hbox{$\sim$}}}}

\begin{document}
\topmargin -0.5cm
\oddsidemargin -0.3cm
\evensidemargin -0.8cm
\pagestyle{empty}
\begin{flushright}
{CERN-TH.7427/94}
\end{flushright}
\vspace*{5mm}
\begin{center}
{\bf HIGHER-ORDER QCD CORRECTIONS}\\{\bf TO DEEP-INELASTIC
SUM RULES}\\
\vspace*{1cm}
{\bf Andrei L. Kataev}\\
%\footnote{On leave of absence from Institute for
%Nuclear Research of the Russian Academy of Sciences, 117312 Moscow,
%Russia} \\
\vspace{0.3cm}
Theoretical Physics Division, CERN,
CH-1211 Geneva 23, Switzerland; \\
Institute for Nuclear Research of the Russian Academy of Sciences,\\
117312 Moscow, Russia.\\
\vspace*{1cm}
{\bf ABSTRACT} \\
\end{center}
\vspace*{2mm}
\noindent
The brief review of the current status of the studies of the effects
of the higher-order perturbative QCD corrections to the deep-inelastic
sum rules is presented.

\vspace*{4cm}
\noindent{Invited talk at the 27th International Conference on
High Energy Physics, Glasgow, Scotland, UK,
21-27 July 1994, shortened version to be published in the Proceedings}
%\rule[.1in]{16.5cm}{.002in}

\begin{flushleft} CERN-TH.7427/94 \\
September 1994
\end{flushleft}

%\thispagestyle{empty}
%\vfill\eject
%\pagestyle{empty}
%\clearpage\mbox{}\clearpage

\newpage
\setcounter{page}{1}
\pagestyle{plain}

{\bf 1.  Introduction}

Up to recently the consideration of the
QCD predictions for the structure functions (SFs) of  deep-inelastic
scattering (DIS) was the basic sourse of  information about the
structure of a nucleon. However, the recent measurements
of the SFs of both polarized and non-polarized DIS \cite{g1},\cite{CCFR}
in the wide interval of the $x=Q^2/2pq$ variable open the possibility
of a more precise determination of the number of the DIS sum
rules (SRs), namely of the polarized Bjorken SR $BjpSR
=\int_0^1 g_1^{ep-en}(x,Q^2)dx$,
 the polarized Ellis-Jaffe SR $EJSR
 = \int_0^1 g_1^{p(n)}(x,Q^2)dx$
 and of the non-polarized Gross-Llewellyn Smith SR
$GLSSR=(1/2)\int_0^1F_3^{\nu p+\overline{\nu}p}(x,Q^2)dx$.
In view of this experimental
progress the detailed studies of the theoretical predictions
for the DIS SRs started to attract special attention. In this talk
we  concentrate on the discussions of the effects of the
perturbative QCD corrections to these quantities.

{\bf 2. The polarized Bjorken SR}

The theoretical expression for the $BjpSR$ has the following form
\begin{equation}
BjpSR=
\frac{1}{3}\mid\frac{g_A}{g_V}\mid \,\bigg[1-a(1+\sum_{i\geq 1}\,
d_{i}a^{i})+ O(\frac{1}{Q^2})\bigg] ,
\label{1}
\end{equation}
where $a=\alpha_s/\pi$ and the exact expressions for the
coefficients $d_1$ and $d_2$, namely $d_1^{ex}=4.583-0.333f$
and $d_2^{ex}=41.440-7.607f+0.177f^2$, were calculated
in the $\overline{MS}$ scheme in Refs. \cite{GL} and \cite{LV}
respectively. However, this scheme is not the unique prescription
for fixing the renormalization scheme ambiguities. For example,
one can use the principle of minimal sensitivity (PMS) \cite{PMS}
or the effective charges (ECH) approach \cite{ECH}. These methods
assume  the role of  ``optimal'' prescriptions, in the sense
that they might provide better convergence of the corresponding
approximations for physical quantities in the non-asymptotic regime.
Therefore, applying these methods, one can try to estimate the
effects of the  $O(a^{N+1})$ corrections starting from the $N$-th
order approximation $D_N^{opt}(a_{opt})$.
 As was originally explained
 in Ref. \cite{PMS}, rewriting the $N$-th order optimized
  expression $D_N^{opt}(a_{opt})$ of the physical quantity
  in terms of the coupling constant $a$ of the
initial scheme one can get the following relation
$D_N^{opt}(a_{opt})=D_N(a)+\delta D_N^{opt} a^{N+1}$. It is now
possible
 to consider the term $\delta D_N^{opt}$ as the one, that simulates
the coefficient of the $O(a^{N+1})$ correction to the
physical quantity $D(a)$ calculated in the certain initial scheme.
Its concrete form $\delta D_N^{opt}=\Omega_N(d_i,c_i)
-\Omega_N^{opt}(d_i^{opt},c_i^{opt})$ ($1\leq i \leq N-1$)
depends on the coefficients $d_i$
of the physical quantity  and $c_i$ of the QCD
$\beta$-function defined as $\beta(a)
=-\beta_0a^2(1+\sum_{i\geq 1}\, c_i a^i)$. The correction terms
$\Omega_N^{opt}$ reflect the dependence on the way of
realization of the ``optimal'' prescription and are rather small.
Within the ECH approach one has
$\Omega_N^{opt}=0$. Moreover, in the case of the PMS approach,
$\Omega_3^{PMS}=0$ \cite{KS} and therefore $\delta D_3^{ECH}
=\delta D_3^{PMS}$.

The above-mentioned procedure was recently used to estimate the
 $O(a^4)$ correction to the $BjpSR$ \cite{KatSt1,KatSt2} and
to roughly fix the uncertainty in the value of the $O(a^5)$ term
\cite{KatSt2}. The table, taken from Ref. \cite{KatSt2}, summarizes
the results of estimates of the coefficients $d_i^{est}$, obtained
by re-expansion of the ECH  for
the $BjpSR$ into the initial $\overline{MS}$ scheme, and demonstrates
their dependence on the number of flavours $f$.
We consider the satisfactory agreement of the obtained estimates
$d_2^{est}$ with the results $d_2^{ex}$
of Ref. \cite{LV} as an argument in favour of the applicability of
this procedure.
\begin{center}
\begin{tabular}{|c|c|c|c|c|} \hline
$f$ & $d^{ex}_{2}$ & $d^{est}_{2}$ &
$d^{est}_{3}$ & $d_4^{est}-c_3d_1$\\ \hline
1 & 34.01 & 27.25 &  290 & 2557\\ \hline
2 & 26.93 & 23.11 &   203 & 1572 \\ \hline
3 & 20.21 & 19.22 &  130 & 854\\ \hline
4& 13.84 & 15.57 &  68 & 342 \\ \hline
5 & 7.83 & 12.19 &  18 & 27 \\ \hline
6 & 2.17 & 9.08 &  -22 & -135 \\ \hline
\end{tabular}
\end{center}
\vspace*{0.5mm}
 The results of estimates of the $NNLO$, $N^3LO$ and $N^4LO$
 corrections in the series for $BjpSR$.
\vspace*{0.5mm}

%We consider the satisfactory agreement of the obtained estimates
%$d_2^{est}$ with the results $d_2^{ex}$ of the exact calculations
%of Ref.\cite{LV} as an argument in favour of applicability of
%this procedure.
The  existing ambiguities  in  $d_4^{est}$
are related to the assumption used that the real value of $d_3$ does
not differ from  $d_3^{est}$ and
to the lack of knowledge of the 4-loop coefficient
of the QCD $\beta$-function. However, even without application of
any additional assumption about its value (e.g. for $f=3,4,5$
 one can use
 the ``geometric progression'' guess $c_3=c_2^2/c_1$), it is
possible to conclude that in the currently available region of energies
$Q^2$=2--10 $GeV^2$ the higher-order QCD corrections to the $BjpSR$
are not negligibly small. The results discussed were used in the
process of the determination of the  value of $\alpha_s(M_Z)=0.122
^{+0.005}_{-0.009}$ \cite{EK} using the $BjpSR$ measurements
\cite{g1}. This result should be compared with the
result $\alpha_s(M_Z)=0.115\pm 0.005(exp)\pm 0.003(th)$
extracted in Ref. \cite{ChK} from the $GLSSR$ data \cite{CCFR}.
Its theoretical uncertainty
comes from the uncertainty of
the estimates \cite{BrK} of the higher-twist terms.

{\bf 3. The Ellis-Jaffe SR}

The theoretical expression for the
$EJSR$ consists of two parts:
$EJSR(Q^2)=EJ_{NS}(Q^2)+EJ_{SI}(Q^2)$. The first non-singlet part
is a renormalization-group-invariant quantity and, apart from the
overall factor, coincides with $BjpSR$. For the case of $f=3$ active
flavours, one has \cite{KatSt1,KatSt2}:
\begin{equation}
%\label{ns}
EJ_{NS}^{p(n)}=\bigg[ 1-a-3.583a^2 - 20.215a^3  - 130
a^4-O(a^5)\bigg]
\times \left( \pm  a_3/12 +  a_8/36 \right)+O\bigg(\frac{1}{Q^2}\bigg),
\label{ns}
\end{equation}
where $a_3=\Delta u-\Delta d$, $a_8=\Delta u+\Delta d-
2\Delta s$ and $\Delta u$, $\Delta d$, $\Delta s$ can be interpreted
as the measure of the polarization of the quarks in a nucleon.
%The order $O(a^4)$ contribution is the result of the estimates
%of Refs.\cite{KatSt1,KatSt2} (see the Table).
%The singlet
%part recieves the contribution from the anomalous dimension
%factor
%$exp(\int_0^{a(Q^2)}\gamma_{SI}(x)/\beta(x)dx)$.
%The corresponding anomalous dimension starts its expansion
%from the order $O(a^2)$ level, namely $\gamma(a)=\sum_{i\geq 1}
%\gamma_ia^{i+1}$.
The  $O(a^2)$ correction to $EJ_{SI}$, which contains the
 anomalous-dimension term, was
calculated recently \cite{Lar}. In order to estimate
the value of the $O(a^3)$ correction the methods of Ref. \cite{PMS}
were supplemented by the considerations of the quantities with
anomalous dimensions \cite{Kat1}. For $f=3$ numbers of flavours,
the estimates were obtained \cite{Kat2}
for the renormalization-invariant
definition of the singlet contribution
and in the case when the $Q^2$-dependence of $\Delta\Sigma=
\Delta u+\Delta d+\Delta s$ is specified.
The  more definite renormalization-invariant estimates have the
following form \cite{Kat2}:
\begin{equation}
EJ_{SI}=\bigg[ 1-0.333a - 0.549a^2 - 2a^3
\bigg] {1\over 9}
\Delta\Sigma_{inv} +O\bigg(\frac{1}{Q^2}\bigg).
\label{19}
\end{equation}
In order to obtain the
 $O(\alpha_s^3)$  estimates
in the case when the $Q^2$-dependence of $\Delta\Sigma$=
$\Delta u$+$\Delta d$+$\Delta s$ is specified,
it is necessary
to fix the value of the 4-loop coefficient of the
 corresponding anomalous dimension function, which starts its expansion
from the  $O(a^2)$ level, namely $\gamma(a)=\sum_{i\geq 1}
\gamma_ia^{i+1}$ \cite{Kat2}. For the case of $f=3$ the final
result \cite{Kat2} reads:
\begin{equation}
EJ_{SI}=
\bigg[ 1-a-1.096a^2 - 3.7a^3 \bigg] {1\over 9}
\Delta\Sigma (Q^2)+O(\frac{1}{Q^2})~.
\label{20}
\end{equation}
Note that we used an additional assumption
 about the value of the
non-calculated term $\gamma_3$  \cite{Kat2}: $\gamma_3\approx
\gamma_2^2/\gamma_1$, whereas the expression
for the $\gamma_2$-term is known \cite{and}.

It can be seen that the perturbative contributions to Eqs. (3),(4)
  are
 significantly smaller than the coefficients of Eq. (2).
 This fact has
an important phenomenological consequence, namely the possibility
of describing available  data of Ref.\cite{g1} by allowing
$\Delta s$ to be non-zero \cite{EK}. The  outcomes of
the fits  \cite{EK} are: $\Delta s=-0.10\pm 0.03$, $\Delta\Sigma
=0.31\pm0.07$ at $Q^2=10\ GeV^2$. Note, however,
that in view of the controversial claims about the possible
contributions of the higher-twist terms in Eqs. (\ref{ns})-(\ref{20})
\cite{ht}, the  analysis of the polarized DIS data
\cite{g1} deserves further experimental and theoretical studies.
One of the possible  clarifying advancements
 could be the determination, from the experimental
data, of the $Q^2$-dependence of the $BjpSR$
and of the $EJSR$. In the case of the $GLSSR$ this work was already
started \cite{KatSid}. Another important theoretical problem is
related to the study of the consequences of the manifestation of
the axial anomaly in the theoretical expression for the $EJSR$
\cite{ET}.

{\bf 4.  DIS vs. $e^+e^-$ annihilation}

The new non-trivial connection between the
 characteristics of the $e^+e^-$ annihilation and DIS
was discovered recently \cite{BK}:
\begin{equation}
D^{e^+e^-}(Q^2)\times GLSSR(Q^2)
\sim 1+\frac{\beta(a)}{a}\bigg[C_1a+C_2a^2\bigg]
+O(a^4)
\label{rel}
\end{equation}
These characteristics are the analytical $O(a^3)$ approximations
of the function $D^{e^+e^-}(Q^2)$ \cite{GKL,SS} and of the
$GLSSR$ \cite{LV}. In Eq.(5)
 $C_1$ and $C_2$ are the analytical numbers, which depend
on the structure of the non-Abelian gauge group.
It should be stressed that the corresponding
perturbative expression for the
$GLSSR$ equals  the one of the $BjpSR$ plus the  $O(a^3)$
contribution of the light-by-light-type graphs \cite{LV}.
Equation (\ref{rel})
demonstrates the appearance of the radiative corrections in
 quark-parton formula of
Ref. \cite{Cr} starting from the $O(a^2)$ level.
This formula is connecting
the amliyude, related to the axial anomaly,
for the $\pi^0\rightarrow\gamma\gamma$
decay with  the quark-parton expressions
for the $D^{e^+e^-}$ function and the $BjpSR$. The most interesting,
yet non-explained, feature of Eq.(\ref{rel}) is the factorization
of the factor $\beta(a)/a$ at the  $O(a^3)$-level. We believe that
the detailed study of this relation could have important theoretical
and phenomenological consequences.

{\bf Acknowledgements}

 We are
grateful to E. Hughes and W.J. Stirling for their invitation
to present the talk at this Conference.

\end{document}